\documentclass[preprint2]{aastex}
\usepackage{amsmath}
\usepackage{graphicx}
\begin{document}
\title{A Method for Smooth Merging of Electron Density Distributions at the Chromosphere-Corona Boundary}

\author{Leonid Benkevitch and Divya Oberoi}
\affil{MIT Haystack Observatory, Westford MA}

\begin{abstract} 
The electron number density ($N_e$) distributions in solar chromosphere and corona are usually described with models of different nature: exponential for the former and inverse power law for the latter. 
Moreover, the model functions often have different dimensionality, e.g. the chromospheric distribution may depend solely on solar altitude, while the coronal number density may be a function of both altitude and latitude. 
For applications which need to consider both chromospheric and coronal models, the chromosphere-corona boundary, where these functions have different values as well as gradients, can lead to numerical problems.
We encountered this problem in context of ray tracing through the corona at low radio frequencies, as a part of effort to prepare for the analysis of solar images from new generation radio arrays like the Murchison Widefield Array (MWA), Low Frequency Array (LOFAR) and Long Wavelength Array (LWA).
We have developed a solution to this problem by using a {\em patch} function, a thin layer between the chromosphere and the corona which matches the values and gradients of the two regions at their respective interfaces.
We describe the method we have developed for defining this patch function to seamlessly ``stitch'' chromospheric and coronal electron density distributions, and generalize the approach to work for any arbitrary distributions of different dimensionality. We show that the complexity of the patch function is independent of the stitched functions dimensionalities. It always has eight parameters (even four for univariate functions) and they may be found without linear system solution for every point. The developed method can potentially be useful for other applications.\\
\end{abstract}

\section{Introduction}

The visible surface of the sun, the photosphere, has the radius $1 R_{\odot} \approx 6.955 \times 10^5$ km. 
The photosphere is surrounded by the solar atmosphere, which consists of two major layers. 
The lower layer, the chromosphere, is relatively thin. It extends up to several thousand kilometers above the photosphere. 
The upper layer, the corona, is a gaseous envelope with the density rapidly decreasing with the radius. 
Depending upon the nature of the study, a working number for size of the corona can be determined based on the coronal height where the coronal plasma becomes too tenuous to be discernible for intended purposes. 
The size of the corona might thus be considered to range from tens up to hundreds of the solar radii. 

Electromagnetic rays at low frequencies in the heliosphere propagate in the regions with the plasma density lower than the critical density $\rho_{cr}$, defined as
\begin{equation}
  \label{rho_cr}
  \rho_{cr} =  \frac{m_p m_e \omega^2}{4 \pi e^2},
\end{equation}
where $e$ is the electron charge, $m_e$ is the electron mass, $m_p$ is the proton mass, and $\omega$ is the radio wave frequency in rad s$^{-1}$. At the critical plasma density the dielectric permittivity $\epsilon$ of the plasma and, hence, its index of refraction for a given frequency $\omega$ become zero. This implies that for most coronal models, radiation at frequencies $< 150$--$180$ MHz does not penetrate to the chromosphere and travels only in the corona.
At higher frequencies, the rays with sharp angles of incidence can penetrate into the much denser and cooler chromosphere. Modeling of propagation of radio frequency radiation through the solar atmosphere hence requires knowledge of both chromospheric and coronal electron density ($N_e$) distributions.
Our method of choice for ray tracing \citep{ben10}, a fast, second order algorithm, requires smoothness in both $N_e$ and its gradient ($\nabla N_e$) everywhere in the medium.
The fact that chromospheric and coronal density distributions come from independent models which are not consistent at the boundary between them leads to a discontinuity in $N_e$ and $\nabla N_e$ at this boundary.
Though the transition between the chromosphere and the corona is intrinsically very rapid and involves rather large density gradients, the discontinuity in $N_e$ and $\nabla N_e$ is unphysical and leads to numerical instability in the ray tracing algorithm.
It is therefore essential to make these distributions consistent at their interface and get them to smoothly morph from one to another while maintaining a smooth gradient.
An attendant problem is the different dimensionality of the chromospheric and coronal density distributions: the former is one-dimensional, depending only on the solar radius, while the latter can be two-dimensional, depending both on the radius and (co)latitude.

In this work we offer our solution to this problem. We use the concept of a thin {\em patch} region which encloses the chromosphere/corona interface, and which is consistent with the chromospheric and coronal models in either region. The {\em patch} is a function constructed to allow the two regions to smoothly morph into each other and accommodates the differences in their dimensionalities.
In Section 2 we present the plasma density models and a mathematical formulation of the problem. 
In Section 3 we start with the simplest case of stitching two univariate functions. 
The method is developed further for the case of merging of a univariate and a multivariate functions in Section 4. 
The method appears to be universal enough to extend it to the problem of merging two multivariate functions along one dimension. 
The appropriate theory is developed in Section 5. 
Numerical implementation details are discussed in Section 6, in particular it is shown that the suggested merging method is not computationally intensive. 

We came across this problem while building the formalism to implement ray tracing through the solar atmosphere at low radio frequencies.
Our work is motivated by the expected near term availability of high fidelity solar images from the new generation radio arrays like the Murchison Widefield Array (MWA) \citep{lon09}.
The early results from the 32 element MWA engineering prototype are very promising and currently represent the state of the art in high fidelity solar imaging at these frequencies \citep{obe11}.
This work is of direct relevance for pursuing solar science with other new generation arrays like the Low Frequency Array (LOFAR) \citep{dev09} and Long Wavelength Array (LWA) \citep{ell09}.
We hope this formulation is general enough to be applicable to similar problems in other independent contexts.

\section{Formulation}
There are a number of models describing the
electron number density $N_e(\mbox{cm}^{-3})$ in the chromosphere and the corona.
For the chromosphere we use the model by Cillie and Menzel \citep{cil35}:
\begin{equation}
\label{Menzel}
N_e(r) = 5.7 \times 10^{11} e^{-7.7 \times 10^{-4}(R_{\odot}(r-1) - 500)},
\end{equation}
where $r$ is the distance from the center of sun measured in the solar radii, $R_{\odot}$. 
This model is applicable for a chromosphere with a thickness of about 10,000 km, or $0.014378R_{\odot}$. 
The simplest available coronal models are one dimensional in nature, depending only on the solar distance, e.g. the Newkirk model \citep{new61}:
\begin{equation}
\label{Newkirk}
N_e(r) = 4.2 \times 10^{4+4.32/r},
\end{equation}
and the Baumbach and Allen model \citep{all47}:
\begin{equation}
\label{Baumbach}
N_e(r) = 10^8(1.55r^{-6} + 2.99r^{-16}).
\end{equation}
where $r$ is the solar distance in units of $R_{\odot}$. 

However, observations show that the plasma density in the corona falls faster the solar distance in the polar regions than it does in equatorial region, especially close to solar minima.
This gave rise to development of more complex models. 
A popular one is by Kuniji Saito \citep{sai70}. This coronal density model depends on two variables, the spherical coordinate $r$ and $\theta$, representing the radial distance from the center of the Sun and the colatitude, respectively:
\begin{align}
\label{Saito}
N_e(r,\theta) &= 3.09 \times 10^8 r^{-16} (1 - 0.5\cos \theta) + \nonumber \\
              & \quad 1.56 \times 10^8 r^{-6} (1 - 0.95\cos \theta) + \nonumber \\ 
              & \quad 0.0251 \times 10^8 r^{-2.5} (1 - \sqrt{\cos \theta}).
\end{align}
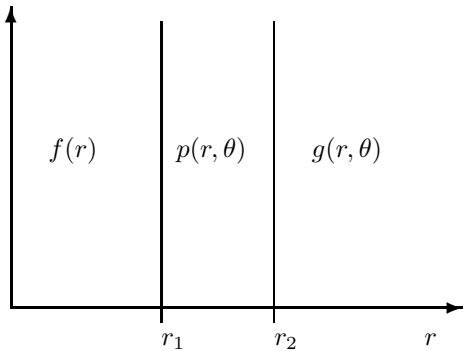
\begin{figure}[!htpb]
\setlength{\unitlength}{1.00mm}
\centering
\begin{picture}(80,60)
  \thicklines
  \put(10,10){\vector(1,0){60}}
  \put(10,10){\vector(0,1){40}}
  \thinlines
  \put(30,8){\line(0,1){40}}
  \put(45,8){\line(0,1){40}}
  \put(65,5){$r$}
  \put(15,30){$f(r)$}
  \put(32,30){$p(r,\theta)$}
  \put(50,30){$g(r,\theta)$}
  \put(30,5){$r_1$}
  \put(45,5){$r_2$}
\end{picture}
\caption{\small Two functions $f(r)$ and $g(r,\theta)$, defined over different parts of space, are stitched together by a third, patch function $p(r,\theta)$ in the intermediate space segment between
  $r=r_1$ and $r=r_2$.\label{Fig1}}
\end{figure} 

Consider a problem of stitching together the chromospheric and coronal models described in Eqs. \ref{Newkirk} and \ref{Saito}, respectively.
Consider a relatively thin transition layer extending from $r_1$ to $r_2$. where $r_1$ is at the top of the chromosphere, and $r_2$ is at the bottom of the corona. 
For this specific problem the limits can be chosen as $r_1=9,000$ km and $r_2=11,000$ km above the photosphere. 
This transition layer encloses the chromosphere-corona boundary. 
Denote the $N_e$ distribution functions in the chromosphere and the corona by  $f(r)$ and $g(r,\theta)$, respectively, as shown in Fig. \ref{Fig1}. 
The {\em patch} function, $p(r,\theta)$, defined in the transition layer between $r_1$ and $r_2$, must match $f(r)$ and $g(r,\theta)$ at both its ends in both its value and derivatives. 
Finding an appropriate $p(r,\theta)$ requires (1) choosing a family of functions for $p(r,\theta)$ and (2) finding its parameters to satisfy the boundary conditions for smoothness.

Note that $f(r)$ and $g(r,\theta)$ have different number of independent variables. Our approach can adequately accommodate this complication, though we start by considering a simpler case of stitching two one-dimensional distributions.
  
\section{Merging one-variable functions}

Consider the case where both chromospheric and coronal electron number density distributions are only dependent on one variable, the radial distance $r$. As shown schematically in Fig.~\ref{Fig1}, $r_1 < r_2$, $f(r)$ on the left is the Cillie and Menzel distribution \eqref{Menzel}, and $g(r)$ on the right is either Newkirk or Baumbach distribution. 
We want the values of $p(r)$ at $r_1$ and $r_2$ to be equal to those of the distributions on the left and right ends, respectively. 
Also, we want the derivative of $p(r)$, denoted as $p'(r)$, to take boundary values equal those of the derivatives of the distributions, $f'(r)$ and $g'(r)$ on the left and on the right, respectively. 
Putting the boundary conditions together in one system:
\begin{equation}
  \label{Sys1D}  
  \left\{
    \begin{aligned}
      p(r_1) &= f(r_1)\\
      p(r_2) &= g(r_2)\\
      p'(r_1) &= f'(r_1)\\
      p'(r_2) &= g'(r_2)\\
    \end{aligned}
  \right.
\end{equation}

We have four equations, therefore $p(r)$ must have exactly four unknown parameters to be fully determined through solving system \eqref{Sys1D}. We also would like to have the simplest system to solve, i.e. a linear system. These considerations lead to the choice of a third order polynomial as $p(r)$,
\begin{equation}
\label{Poly3ord}
p(r) = a_0 + a_1r + a_2r^2 + a_3r^3.
\end{equation}
Substitution of the polynomial \eqref{Poly3ord} into the system \eqref{Sys1D} yields the linear system
\begin{equation}
  \label{Sys1Dpoly}  
  \left\{
    \begin{aligned}
      a_0 + a_1r_1 + a_2r_1^2 + a_3r_1^3 &= f(r_1)\\
      a_0 + a_1r_2 + a_2r_2^2 + a_3r_2^3 &= g(r_2)\\
      a_1 + 2a_2r_1 + 3a_3r_1^2 &= f'(r_1)\\
      a_1 + 2a_2r_2 + 3a_3r_2^2 &= g'(r_2)
    \end{aligned}
  \right.
\end{equation}
Its solution, the vector $\mathbf a = 
\begin{Vmatrix} a_0; & a_1; & a_2; & a_3 \end{Vmatrix}^T$, comprises the coefficients of polynomial \eqref{Poly3ord}. 
Once these coefficients have been determined, the patch polynomial \eqref{Poly3ord}, by construction, has value and derivatives equal to those of the chromospheric distribution $f(r)$ at its left end and the coronal distribution $g(r)$ at its right. 
\begin{figure}[!htpb]
\plotone{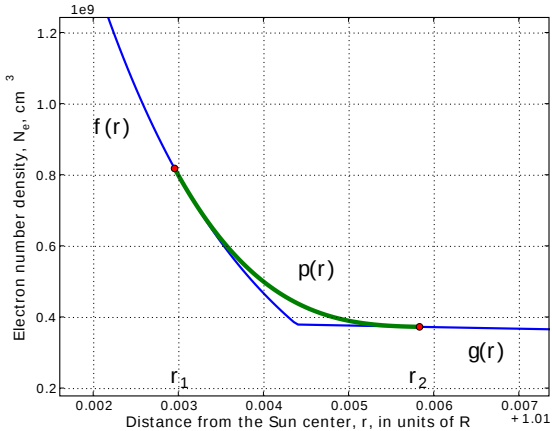}
\caption{\small Stitching of the two functions $f(r)$ 
and $g(r)$, by a patch function $p(r)$  between  $r=r_1$ and $r=r_2$.
\label{Fig2}}
\end{figure} 
The system \eqref{Sys1Dpoly} can be also written compactly in the matrix-vector form,
\begin{equation}
  \label{Sys1DpolyMxVec}
  \mathbf P \mathbf a = \mathbf c,
\end{equation}
where $\mathbf c =
\begin{Vmatrix} f(r_1); & g(r_2); & f'(r_1); & g'(r_2) \end{Vmatrix}^T$ is
the right-hand-side vector, and $\mathbf P$ is the system matrix,
\begin{equation}
  \label{MatrixP}
  \mathbf{P} = 
  \begin{pmatrix} 
    1 & r_1 & r_1^2 & r_1^3  \\
    1 & r_2 & r_2^2 & r_2^3  \\
    0 & 1   & 2r_1  & 3r_1^2 \\
    0 & 1   & 2r_2  & 3r_2^2
  \end{pmatrix}
\end{equation}

An example of smooth stitching of the chromospheric and coronal $N_e$ distributions, given by equations \eqref{Menzel} and \eqref{Baumbach}, is given in Fig. \ref{Fig2}. 

\section{Merging one- and multi-variable functions along one dimension}

We now generalize the solution to the case with a function of one variable at $r_1$ and a function of two or more variables at $r_2$. A real-world example here is the problem of merging the model of Cillie and Menzel \citep{cil35}, Eq.~\eqref{Menzel} in the chromosphere with the model of \citet{sai70}, Eq.~\eqref{Saito} in the corona. The patch $p(r,\theta)$ is now a function of two variables. It must smoothly match both values and derivatives of $f(r)$ at its left boundary and of $g(r, \theta)$ at the right boundary. Note that $f(r)$ has only one derivative, while $g(r, \theta)$ has two partial derivatives. In order to guess the kind of a patch function we count the number of constraints (or degrees of freedom) imposed on the patch: 
two on the end point values and four on the partial derivatives at the end points, or six on total. For a completely asymmetric coronal distribution $g(r,\theta,\varphi)$, when it is a function of all the three variables, the radial distance $r$, the colatitude $\theta$, and the longitude $\varphi$, the number of conditions imposed on the patch $p(r,\theta,\varphi)$ becomes $2+3*2=8$. These calculations suggest that the complexity of the patch function apparently grows with the dimensionality of the stitched functions. However, this is not the case. As we shall show here and in the next Section, the patch function can be build such that it always depends only on eight parameters independent of the dimensionality of the stitched functions. The method can be used for stitching functions of arbitrary (and different) numbers of dimensions, along one of them. 

Note that the patch function must gradually turn from
multi-variable $g(r,\theta,\varphi)$ at $r_2$ into single-variable
$f(r)$ at $r_1$. Consider a function $g(r_2,\theta,\varphi)$. It has
its variable $r$ ``frozen'' at $r_2$, so that 
$g(r_2,\theta,\varphi)$ is now dependent only on the two remaining 
variables, $\theta$ and $\varphi$. 
Form the product of the  $g(r_2,\theta,\varphi)$ and some
polynomial $b(r)$, which has value of 0 at $r_1$ and 1 at
$r_2$. 
The function $b(r)g(r_2,\theta,\varphi)$  is not only equal to 0
at $r_1$, its partial derivatives with respect to $\theta$ and $\varphi$ 
also evaluate to 0 at $r_1$, i.e. at $r_1$ it effectively depends on $r$ only. 
However, it cannot be linked to $f(r)$ directly because
generally $f(r_1) \ne 0$. To make such linkage possible, we add one more
polynomial, $a(r)$, so the patch function takes the form 
\begin{equation}
  \label{PatchFun}
  p(r,\theta,\varphi) = a(r) + b(r)g(r_2,\theta,\varphi).
\end{equation}

The polynomial $a(r)$ must be equal $f(r)$ at $r_1$, and must be equal zero at $r_2$, 
where $b(r_2)g(r_2,\theta,\varphi)$ equals $g(r,\theta,\varphi)$. Thus, the conditions imposed on the polynpmials are
\begin{equation}
  \label{BndCondab}
  \begin{aligned}
    &a(r_1) = f(r_1); &a(r_2) = 0; \\
    &b(r_1) = 0; &b(r_2) = 1.
  \end{aligned}
\end{equation}
For the multi-variable functions the boundary conditions include their partial derivatives. Listing the boundary conditions as a system of equations yields:
\begin{equation}
  \label{BoundaryCondMultivar}  
  \left\{
    \begin{aligned}
      p(r_1,\theta,\varphi) &= a(r_1) + b(r_1)g(r_2,\theta,\varphi) =
        f(r_1) \\
      p_r(r_1,\theta,\varphi) &= a_r(r_1) + b_r(r_1)g(r_2,\theta,\varphi) =
        f_r(r_1) \\
      p_{\theta}(r_1,\theta,\varphi) &= 
        b(r_1)g_{\theta}(r_2,\theta,\varphi) = 0 \\
      p_{\varphi}(r_1,\theta,\varphi) &=
        b(r_1)g_{\varphi}(r_2,\theta,\varphi) = 0 \\
      p(r_2,\theta,\varphi) &= a(r_2) + b(r_2)g(r_2,\theta,\varphi) \\
        &= g(r_2,\theta,\varphi) \\
      p_r(r_2,\theta,\varphi) &= a_r(r_2) + b_r(r_2)g(r_2,\theta,\varphi) \\
        &= g_r(r_2,\theta,\varphi) \\
      p_{\theta}(r_2,\theta,\varphi) &=
        b(r_2)g_{\theta}(r_2,\theta,\varphi) = g_{\theta}(r_2,\theta,\varphi) \\
      p_{\varphi}(r_2,\theta,\varphi) &=
        b(r_2)g_{\varphi}(r_2,\theta,\varphi) = g_{\varphi}(r_2,\theta,\varphi) 
      \end{aligned}
  \right.
\end{equation}
Here we denote a partial derivative with respect to a
variable by putting this variable at the subscript position after the
function name, e.g. $g_{r}(r,\theta,\varphi)$ is the
derivative of $g(r,\theta,\varphi)$ with respect to $r$. 
Now we simplify the system \eqref{BoundaryCondMultivar} using conditions \eqref{BndCondab} on the polynomials $a(r)$ and $b(r)$:
\begin{equation}
  \label{BdCondMvarSimplified}  
  \left\{
    \begin{aligned}
      p(r_1,\theta,\varphi) &= a(r_1) = f(r_1) \\
      p_r(r_1,\theta,\varphi) &= a_r(r_1) + b_r(r_1)g(r_2,\theta,\varphi) =
        f_r(r_1) \\
      p_{\theta}(r_1,\theta,\varphi) &= 0 = 0 \\
      p_{\varphi}(r_1,\theta,\varphi) &= 0 = 0 \\
      p(r_2,\theta,\varphi) &= g(r_2,\theta,\varphi) = g(r_2,\theta,\varphi) \\
      p_r(r_2,\theta,\varphi) &= a_r(r_2) + b_r(r_2)g(r_2,\theta,\varphi) \\
        &= g_r(r_2,\theta,\varphi) \\
      p_{\theta}(r_2,\theta,\varphi) &=
        g_{\theta}(r_2,\theta,\varphi) = g_{\theta}(r_2,\theta,\varphi) \\
      p_{\varphi}(r_2,\theta,\varphi) &= 
        g_{\varphi}(r_2,\theta,\varphi) = g_{\varphi}(r_2,\theta,\varphi) 
      \end{aligned}
  \right.
\end{equation}
We can see that the first, the third, the fourth, the fifth, the seventh, and the
eighth equations in \eqref{BdCondMvarSimplified} are nothing but
tautologies and can be removed from the system, which now takes the
form
\begin{equation}
  \label{BdCondMvarSimpl2}  
  \left\{
    \begin{aligned}
      a_r(r_1) + b_r(r_1)g(r_2,\theta,\varphi) &= f_r(r_1) \\
        &= p_r(r_1,\theta,\varphi) \\
      a_r(r_2) + b_r(r_2)g(r_2,\theta,\varphi) &= g_r(r_2,\theta,\varphi) \\
        & = p_r(r_2,\theta,\varphi) 
    \end{aligned}
  \right.
\end{equation}
If we for simplicity fix the $a(r)$ derivatives at zero, this set of equations provides boundary conditions for the derivatives of $b(r)$, :
\begin{equation}
  \label{CondsOnDerivab}  
  \left\{
    \begin{aligned}
      a_r(r_1) &= 0 \\
      a_r(r_2) &= 0 \\
      b_r(r_1) &=  \frac{f_r(r_1)}{g(r_2,\theta,\varphi)} \\
      b_r(r_2) &= \frac{g_r(r_2,\theta,\varphi)}{g(r_2,\theta,\varphi)}.
      \end{aligned}
  \right.
\end{equation}
We remember that there are four constraints imposed on each of the polynomials: two on their end-point values and two on their partial derivatives with respect to the stitching-dimension variable $r$. Therefore, both polynomials are of the $3^{rd}$ order, have four unknown coefficients each, and require eight equations to be solved for. We have four equations in \eqref{CondsOnDerivab}; the other four are in \eqref{BndCondab}. Combining the both provides two linear systems that can be used to determine coefficients of the polynomials $a(r)$ and $b(r)$: 

\begin{equation}
  \label{CondsOnDeriva}  
  \left\{
    \begin{aligned}
      a(r_1) &= f(r_1) \\
      a(r_2) &= 0 \\
      a_r(r_1) &= 0 \\
      a_r(r_2) &= 0 
      \end{aligned}
  \right.
\end{equation}
and
\begin{equation}
  \label{CondsOnDerivb}  
  \left\{
    \begin{aligned}
      b(r_1) &= 0 \\
      b(r_2) &= 1 \\
      b_r(r_1) &=  \frac{f_r(r_1)}{g(r_2,\theta,\varphi)} \\
      b_r(r_2) &= \frac{g_r(r_2,\theta,\varphi)}{g(r_2,\theta,\varphi)}.
      \end{aligned}
  \right.
\end{equation}
One can notice that both linear systems have a nice common property,
they use the same system matrix $\mathbf{P}$ as used in the 
previous Section (see Eq.~\eqref{Sys1Dpoly} through
\eqref{MatrixP}). The systems \eqref{CondsOnDeriva} and 
\eqref{CondsOnDerivb} can be rewritten in the vector-matrix form 
as follows:
\begin{equation}
  \label{MxVecSysa}
  \mathbf P \mathbf a = \mathbf{c}_a,
\end{equation}
\begin{equation}
  \label{MxVecSysb}
  \mathbf P \mathbf b = \mathbf{c}_b,
\end{equation}
where $\mathbf a = \begin{Vmatrix} a_0; & a_1; & a_2; &
  a_3 \end{Vmatrix}^T$ and  $\mathbf b = \begin{Vmatrix} b_0; & b_1; & b_2; &
  b_3 \end{Vmatrix}^T$ are vectors of coefficients for the polynomials
\begin{equation}
\label{Polya}
a(r) = a_0 + a_1r + a_2r^2 + a_3r^3.
\end{equation}
\begin{equation}
\label{Polyb}
b(r) = b_0 + b_1r + b_2r^2 + b_3r^3.
\end{equation}
\begin{figure}[!hpt]
\epsscale{1.0}
\plotone{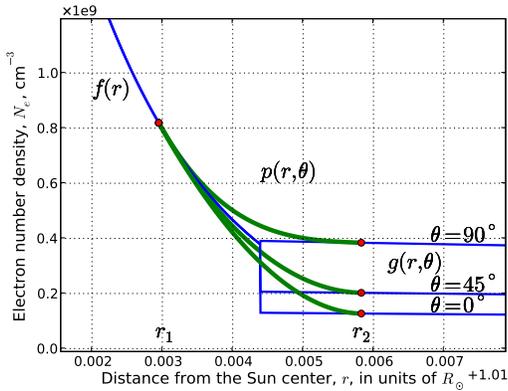}
\caption{\small Stitching of the $f(r)$ function of a single 
variable and the function $g(r,\theta)$ of two independent variables 
along the $r$ dimension between $r=r_1$ and $r=r_2$. The patch function
$p(r,\theta)$ has the form given by Eq.~\eqref{PatchFun}. 
The $g(r,\theta)$ function is plotted for three values of colatitude $\theta$,
$0^{\circ}$, $45^{\circ}$, and $90^{\circ}$. \label{Fig3}}
\end{figure} 
The right-hand-side vectors $\mathbf c_a$ and $\mathbf c_b$ are
determined according to \eqref{CondsOnDeriva} and
\eqref{CondsOnDerivb}:
\begin{equation}
  \label{VecCa}
  \mathbf{c}_a = \parallel
  \begin{array}{cccc}
    f(r_1); & 0; & 0; &  0 
 \end{array} \parallel ^T,
\end{equation}
\begin{equation}
  \label{VecCb}
  \mathbf{c}_b = \parallel
  \begin{array}{cccc}
    0; & 1; &
    \frac{f_r(r_1)}{g(r_2,\theta,\varphi)}; &
    \frac{g_r(r_2,\theta,\varphi)}{g(r_2,\theta,\varphi)}
  \end{array} \parallel ^T.
\end{equation}

Substitution of $a(r)$ and $b(r)$ in \eqref{PatchFun} gives a
smooth patch function to link together the $f()$ and $g()$ functions.

In order to test the method we stitch two distributions dependent on
different number of variables, those by Cillie and Menzel's \citep{cil35}, 
Eq.~\eqref{Menzel} and by \citet{sai70}, Eq.~\eqref{Saito}. 
The former depends only on $r$, while the latter depends
on $r$ and $\theta$. 
The stitching occurs along the $r$ coordinate. Fig. \ref{Fig3} shows three
stitching cases for different colatitudes $\theta$. 
This example illustrates the application of the general method described here. 

\section{Merging two multivariable functions along one dimension}

It is possible to generalize the one-and-multi-variable merging method
to stitch together two multi-variable functions along the axis of one of the
variables they have in common. 
For example, we may need to smoothly merge two
functions, $f(r,\theta,\varphi)$ and $g(r,\theta,\varphi)$ along the
$r$ dimension using the patch function $p(r,\theta,\varphi)$ over the
interval $[r_1,r_2]$. 
The patch function must have the properties of
$f(r,\theta,\varphi)$ (value and derivatives) at the $r_1$ end, 
but on the way from $r_1$ to $r_2$ these properties must smoothly
morph into being indistinguishable from $g(r,\theta,\varphi)$.
To build such a function, consider two functions formed from $f$
and $g$ by ``freezing'' the $r$ variable:  $f(r_1,\theta,\varphi)$ 
and $g(r_2,\theta,\varphi)$. 
Following the method developed in the previous Section, we introduce two polynomials, $a(r)$ and $b(r)$. In the interval $[r_1,r_2]$, $a(r)$ should drop off from one to zero, while $b(r)$ should increase from zero to one. Then the function:
\begin{equation}
  \label{PatchFunMul}
  p(r,\theta,\varphi) = a(r)f(r_1,\theta,\varphi) + b(r)g(r_2,\theta,\varphi)
\end{equation}
will behave as desired to serve as a valid patch function. The boundary conditions imposed on the values of the polynomials $a(r)$ and $b(r)$ are:
\begin{equation}
  \label{BndCondabMul}
  \begin{aligned}
    &a(r_1) = 1; &a(r_2) = 0; \\
    &b(r_1) = 0; &b(r_2) = 1.
  \end{aligned}
\end{equation}
As before, we build a system of equations comprising all the boundary 
conditions on $p(r,\theta,\varphi)$ at the end points $r_1$ and $r_2$ 
similar to system \eqref{BoundaryCondMultivar}:
the values and derivatives of $p(r,\theta,\varphi)$ are equated to those of
$f(r,\theta,\varphi)$ on the left and those of  $g(r,\theta,\varphi)$ on the right. 
After removing the tautologies only two equations remain:
\begin{equation}
  \label{BdCondMvarSimpl2Mul}  
  \left\{
    \begin{aligned}
        a_r(r_1)f(r_1,\theta,\varphi) + b_r(r_1)g(r_2,\theta,\varphi) &=  
        f_r(r_1,\theta,\varphi)\\
        &= p_r(r_1,\theta,\varphi) \\
        a_r(r_2)f(r_1,\theta,\varphi) + b_r(r_2)g(r_2,\theta,\varphi) &= 
        g_r(r_2,\theta,\varphi) \\
        &= p_r(r_2,\theta,\varphi) 
      \end{aligned}
  \right.
\end{equation}
This system of two equations has four unknowns, the derivatives of
polynomials $a(r)$ and $b(r)$ at both ends of the interval
$[r_1,r_2]$. For simplicity, we assign values of zero to $a_r(r_2)$
and  $b_r(r_1)$ to form the system
\begin{equation}
  \label{CondsOnDerivabMul}  
  \left\{
    \begin{aligned}
      a_r(r_1) &=  \frac{f_r(r_1,\theta,\varphi)}{f(r_1,\theta,\varphi)} \\
      a_r(r_2) &= 0 \\
      b_r(r_1) &= 0 \\
      b_r(r_2) &= \frac{g_r(r_2,\theta,\varphi)}{g(r_2,\theta,\varphi)}.
      \end{aligned}
  \right.
\end{equation}
Combining conditions on the polynomials $a(r)$ and $b(r)$ from
\eqref{BndCondabMul} and \eqref{CondsOnDerivabMul} we build two linear
systems:

\begin{equation}
  \label{CondsOnDerivMula}  
  \left\{
    \begin{aligned}
      a(r_1) &= 1 \\
      a(r_2) &= 0 \\
      a_r(r_1) &= \frac{f_r(r_1,\theta,\varphi)}{f(r_1,\theta,\varphi)} \\
      a_r(r_2) &= 0 
      \end{aligned}
  \right.
\end{equation}
and
\begin{equation}
  \label{CondsOnDerivMulb}  
  \left\{
    \begin{aligned}
      b(r_1) &= 0 \\
      b(r_2) &= 1 \\
      b_r(r_1) &= 0 \\
      b_r(r_2) &= \frac{g_r(r_2,\theta,\varphi)}{g(r_2,\theta,\varphi)}
      \end{aligned}
  \right.
\end{equation}
There are four constraints imposed on each of the polynomials, two on the end-point values and two on the partial derivatives with respect to $r$. Both polynomials are cubic, like \eqref{Polya} and \eqref{Polyb}. We already have eight equations to solve for the eight coefficients of $a(r)$ and $b(r)$. Both system have the same system matrix $\mathbf{P}$,
given in \eqref{MatrixP}, and can be compactly written in the
matrix-vector form as shown in \eqref{MxVecSysa} and
\eqref{MxVecSysb}. The right-hand-side vectors for systems
\eqref{CondsOnDerivMula} and \eqref{CondsOnDerivMulb} are 
\begin{equation}
  \label{VecMulCa}
  \mathbf{c}_a = \parallel
  \begin{array}{cccc}
    1; & 0; & \frac{f_r(r_1,\theta,\varphi)}{f(r_1,\theta,\varphi)}; &  0 
 \end{array} \parallel ^T,
\end{equation}
\begin{equation}
  \label{VecMulCb}
  \mathbf{c}_b = \parallel
  \begin{array}{cccc}
    0; & 1; & 0; &
    \frac{g_r(r_2,\theta,\varphi)}{g(r_2,\theta,\varphi)}
  \end{array} \parallel ^T.
\end{equation}

Thus $a(r)$ and $b(r)$ are successfully found. The patch function
\eqref{PatchFunMul} with these polynomials as coefficients smoothly
stitches together the two functions, $f(r,\theta,\varphi)$ and
$g(r,\theta,\varphi)$, along the $r$ dimension on the interval
$[r_1,r_2]$.  Note that the method for stitching of two multivariable
functions is not more complex than the method for stitching a
single-variable function with a multivariable function from the
previous Section. 

So far we considered merging of the functions defined over real,
three-dimensional space with the same set of independent
variables. 
However, using patch function in the form given in
\eqref{PatchFunMul} and the procedure described in this Section it is
easy to show that the merged functions can have arbitrary numbers
of variables, and even different numbers of variables. Having at least one variable in common is the only requirement. 

Consider a more general example. Suppose two functions defined in
eight-dimensional space, $f(x,y,z,q,r)$ and $g(q,r,s,t,u)$ need to be
smoothly merged along the $q$ dimension over the interval
$[q_1,q_2]$. The two functions have different sets of independent variables with non-empty intersection having $q$ as its element. The patch function must be dependent on the union of the both sets of variables.  It is defined in the form
\begin{equation}
  \label{PatchFun8D}
  \begin{aligned}
    p(x,y,z,q,r,t,u,s) &= a(q)f(x,y,z,q_1,r) \\
      &+ b(r)g(q_2,r,s,t,u).
  \end{aligned}
\end{equation}
Following the reasoning from the previous Sections we can show that the cubic polynomials $a(q)$ and $b(q)$ can be determined through solving the linear systems \eqref{MxVecSysa} and \eqref{MxVecSysb} with the right hand side vectors
\begin{equation}
  \label{Vec8DCa}
  \mathbf{c}_a = \parallel 
    \begin{array}{cccc}
    1;  &0;  & \frac{f_q(x,y,z,q_1,r)}{f(x,y,z,q_1,r)};  &0 
 \end{array} \parallel ^T,
\end{equation}
\begin{equation}
  \label{Vec8DCb}
  \mathbf{c}_b = \parallel
  \begin{array}{cccc}
    0; & 1; &  0; & \frac{g_q(q_2,r,s,t,u)}{g(q_2,r,s,t,u)}
  \end{array} \parallel ^T.
\end{equation}

\section{Discussion}

The method of stitching two functions along one dimension 
described here was developed in context of numerical simulations, which need to be
optimized for computational efficiency and speed. 
If both chromospheric and coronal electron
number density distributions depend only on the distance from the sun center $r$, the four coefficients of patch polynomial \eqref{Poly3ord} can be calculated at the 
initialisation step. 
During the simulation there is no need to solve
system \eqref{Sys1Dpoly}; only the polynomial values need to be
calculated. 
For example, in the ray tracing application this only happens 
when the ray enters the transition layer between the chromosphere 
and the corona. 
Therefore, the computational overhead of the stitching
is relatively small.

In more complex cases of spherically non-symmetric coronal
distributions like that by Saito \eqref{Saito}, the linear system
\eqref{MxVecSysb} must be solved 
at every step when the ray happens to travel inside the transition
layer. This is required because the $\mathbf{c}_b$ vector now depends
on the $\theta$ (and probably $\varphi$) variables, whose values may
change from one step to the next.
Here are a few tips to accelerate the computations. 
The fact that both linear systems have
the same system matrix $\mathbf{P}$ \eqref{MatrixP}, is one of the 
benefits of the method. 
$\mathbf{P}$ only depends on the values of $r_1$ and $r_2$ (i.e. boundaries of the transition
layer), which are usually held constant during the course of computations. 
This makes it possible to calculate 
$\mathbf{P}$ once and for all before the actual raytracing starts. To save
even more time one can replace the linear system solving with 
matrix-vector multiplying. 
To do so, $\mathbf{P}^{-1}$, must be calculated during the initialization phase.
The vector $\mathbf{a}$ of polynomial \eqref{Polya} coefficients must 
also be pre-calculated as the product of the matrix $\mathbf{P}^{-1}$ 
and the vector $\mathbf{c}_a$:
\begin{equation}
  \label{PbyCa}
  \mathbf{a} = \mathbf{P}^{-1} \mathbf{c}_a.
\end{equation}
Then, in the process of ray tracing, only the coefficients of the
polynomial $b(r)$ (see \eqref{Polyb}) must be calculated as 
\begin{equation}
  \label{PbyCb}
  \mathbf{b} = \mathbf{P}^{-1} \mathbf{c}_b,
\end{equation}
and followed by computing the patch function \eqref{PatchFun} 
itself. Adopting this scheme makes computations significantly
faster. Note, however, that in case of stitching two three-variable
functions both operations, \eqref{PbyCa} and
\eqref{PbyCa}, must be repeated at every ray tracing step. 

In the previous Section it is shown that this approach to stitching
multi-variate functions is not limited to the three-dimensional cases. 
It is general enough to be applied to abstract functions of any number of independent variables, 
while retaining its simplicity. 
For any number of dimensions only eight parameters, the coefficients of the $3^{rd}$ order 
polynomials  $a(r)$ and $b(r)$ need to be determined.      

\section{Conclusion}
The method developed here is intended for smooth merging of the electron number density model distributions at the boundary between the solar corona and chromosphere. 
Used extensively in our coronal radio propagation studies, the method has shown excellent results. 
We have also shown that the method can easily be generalized to the problems where two abstract functions of arbitrary variable numbers need to be smoothly stitched along one common dimension. 
 
\section*{Acknowledgements}
This work was partially supported under grants from the National Science Foundation, Division of Astronomical Sciences and Atomspheric and Geospace Sciences to the MIT Haystack Observatory.

\end{document}